\documentclass[aps,tightenlines,nofootinbib,preprint,
superscriptaddress,groupedaddress,epsfig]{revtex4}

\usepackage{graphicx}

\begin{document}

\title{Annihilation Rate of Heavy $0^{-+}$ Quarkonium \\
in Relativistic Salpeter Method }
\vspace{2cm}

\author{C. S. Kim}
\email{cskim@yonsei.ac.kr} \affiliation{Department of Physics,
Yonsei University, Seoul 120-749, Korea}
\affiliation{Physics Division, National Center for Theoretical Sciences,
Hsinchu 300, Taiwan}
\author{Taekoon Lee}
\email{tlee@phya.snu.ac.kr} \affiliation{Department of Physics,
Seoul National University, Seoul 151-742, Korea}
\author{Guo-Li Wang}
\email{glwang@cskim.yonsei.ac.kr} \affiliation{Department of
Physics, Yonsei University, Seoul 120-749, Korea}

\baselineskip=20pt

\vspace{2cm}

\begin{abstract}

Two-photon and two-gluon annihilation rates of $\eta_c$,
$\eta'_c$, $\eta_b$ and $\eta'_b$ are estimated in the
relativistic Salpeter method. By solving the full Salpeter
equation with a well defined relativistic wave function, we
estimate $M_{\eta_c}=2.979\pm 0.432$ GeV, $M_{\eta'_c}=3.566\pm
0.437$ GeV, $M_{\eta_b}=9.364\pm 1.120$ GeV and
$M_{\eta'_b}=9.941\pm 1.112$ GeV. We calculated the transition
amplitude using the Mandelstam formalism and estimate the decay
widths: $\Gamma(\eta_c \rightarrow 2\gamma)=7.14\pm 0.95$ KeV,
$\Gamma(\eta'_c \rightarrow 2\gamma)=4.44\pm 0.48$ KeV,
$\Gamma(\eta_b \rightarrow 2\gamma)=0.384\pm 0.047$ KeV and
$\Gamma(\eta'_b \rightarrow 2\gamma)=0.191\pm 0.025$ KeV. We also
give estimates of total widths by the two-gluon decay rates:
$\Gamma_{tot}(\eta_c)=19.6\pm 2.6$ MeV,
$\Gamma_{tot}(\eta'_c)=12.1\pm 1.3$ MeV,
$\Gamma_{tot}(\eta_b)=6.98\pm 0.85$ MeV and
$\Gamma_{tot}(\eta_b)=3.47\pm 0.45$ MeV.

\end{abstract}


\maketitle

\section{Introduction}

It is well known that two-photon or two-gluon annihilation rate of
heavy $0^{-+}$ quarkonium $c\bar c$ or $b \bar b$ is related to
the wave function, so this process will be helpful to understand
the formalism of inter-quark interactions, and can be a
sensitive test of the potential model. With the replacement of
the photons by gluons, the finial state becomes two gluon state, which
will be helpful to give information on the total width of the
corresponding quarkonium.

Experimentally there are quite many  results for the decay width
$\Gamma_{tot}(\eta_c)$ with a wide range of values and
uncertainties by different collaborations;
for example, in recent experiment of
Barbar \cite{barbar} they give 34.3 (2.3) (0.9) MeV, much larger
than the cited value $16.0^{+3.6}_{-3.2}$ MeV by Particle Data
Group \cite{pdg}. However, $\eta'_c$ has been just declared observed by Belle
\cite{belle} and by Barbar \cite{barbar}; $\eta_b$ and $\eta'_b$
have not been observed yet, even though  there were some experiments
to search for $\eta_b$, e.g., the ALEPH
\cite{aleph} collaboration. In short, unlike the
corresponding vector $1^{--}$ quarkonium which can be produced
directly by $e^{+}e^{-}$ annihilation, experiments on
$0^{-+}$ quarkonium have just begun, even for $\eta_c$. This due
to the small cross section; presently there are $57.7\times 10^6$ $J/\Psi$
events collected with the BES-II detector
\cite{fang}, but there are only $2547\pm 90$ $\eta_c$ events
collected by the Barbar detector \cite{barbar}.

For the theoretical estimates of the annihilation rate
for $\eta_c^{(')}$ and $\eta_b^{(')}$, we have various methods
readily available in hand.
First was the non-relativistic calculation, then the
relativistic corrections were found to be important especially for
$c \bar c$ states. In recent years, many authors try to focus on the
relativistic corrections and there are already some versions of
relativistic calculation, and they give improved results over the
non-relativistic methods. In this letter, we give yet another
relativistic calculation by the instantaneous Bethe-Salpeter
method \cite{BS}, which is a full relativistic method
\cite{salp} with a well defined relativistic form of the wave
function.

There are two sources of relativistic corrections; one is the
correction in relativistic kinematics which appears in the decay
amplitudes through a well defined form of relativistic wave
function ($i.e.$ not merely through the wave function at origin);
the other relativistic correction comes via the
relativistic inter-quark dynamics, which requires not only a well
defined relativistic wave function but also a good relativistic
formalism to describe the interactions among quarks.

The Bethe-Salpeter equation is a well-known tool to describe a
relativistic bound state. And the Salpeter equation is the special
case of Bethe-Salpeter equation when the interaction is instantaneous.
It has been shown that the instantaneous approach is a good
approximation in heavy mesons, especially for the equal-mass quarkonium,
since the
non-instantaneous correction  was found to be very small in
equal mass system \cite{changwang2}. The full Salpeter equation
includes two parts of the wave function, the positive  and
negative energy part. In the case of heavy mesons the negative
energy part usually gives a smaller contribution than the
positive energy part, and therefore, to simplify the calculation for the heavy mesons
authors like to make a further approximation to the
Salpeter equation by ignoring the negative part contribution.
However, since we are considering full
relativistic calculation, and the negative energy contribution was
found to be not very small for some cases \cite{munz}, in this letter we
will solve the full Salpeter equation including the negative
contribution, and use the full Salpeter wave
function to estimate the annihilation decay width of quarkonium.

We note that the form of the wave function is also important in the calculation,
since the corrections of the relativistic kinetics come mainly
through it. We begin from the quantum field theory,
analyze the parity and charge conjugation of bound state, and give
a formula for the wave function that is in a relativistic form with
definite parity and charge conjugation symmetry.
Another important thing is how to use the relativistic wave
function of bound state to obtain a relativistic transition
amplitude, since a non-relativistic transition amplitude even with
a relativistic wave function will lose the benefit of
relativistic effects caused by the relativistic wave function.
The Mandelstam formalism is well suited for the computation of
relativistic transition amplitude, and we begin with this formulism
to give a formula of the transition amplitude.

In Sec. II, we give theoretical details for the transition amplitude in
Mandelstam formalism and the corresponding wave function with a well
defined relativistic form. In Sec. III, the full Salpeter equation
is solved, and the mass spectra and numerical value of the wave function
are obtained. Then the two-photon decay width and full width of
heavy $0^{-+}$ quarkonium are estimated. In Sec. III, short
discussions and a summary are also given.

\section{Theoretical Details}

According to the Mandelstam \cite{mandelstam} formalism, the
relativistic transition amplitude of a quarkonium decaying into
two photons (see figure 1) can be written as:
\begin{eqnarray}
T & = & i\sqrt{3}\; (iee_q)^2 \!
 \int \!\! \frac{d^4q}{(2\pi)^4}
              \; \mbox{tr}\; \Bigg\{ \,
    \chi(q) \bigg[\varepsilon\!\!\! /_2\, S(p_1-k_1)\,
    \varepsilon\!\!\! /_1 +
     \varepsilon\!\!\! /_1\, S(p_1-k_2)\,
    \varepsilon\!\!\! /_2 \bigg]
        \Bigg\},
\label{eq1}
\end{eqnarray}
where $k_1$, $k_2$; $\varepsilon_1$, $\varepsilon_2$ are the
momenta and polarization vectors of photons; $e_q=\frac{2}{3}$ for
charm quark and $e_q=\frac{1}{3}$ for bottom quark; $p_1$ and
$p_2$ are the momenta of constitute quark and antiquark;
$\chi(q)$ is the quarkonium Bethe-Salpeter wave function with the
total momentum $P$ and relative momentum $q$, related by
$$p_{_1}={\alpha}_{1}P+q, \;\; {\alpha}_{1}\equiv\frac{m_{1}}{m_{1}+m_{2}}~,$$
$$p_{2}={\alpha}_{2}P-q, \;\; {\alpha}_{2}\equiv\frac{m_{2}}{m_{1}+m_{2}}~.$$

Since $p_{10}+p_{20}=M$, the approximation $p_{10}=p_{20}=\frac{M}{2}$
is a good choice for the equal mass system
\cite{barbieri,keung,chao}. Having this approximation, we can
perform the integration over $q_0$ to reduce the expression,
with the notation for the
Salpeter wave function $\Psi(q)=\int\frac{dq_0}{2\pi}\chi(q)$, to
\begin{eqnarray}
T & = & \sqrt{3}\; (ee_q)^2 \!
 \int \!\! \frac{d\vec{q}}{(2\pi)^3}
              \; \mbox{tr}\; \Bigg\{ \,
    \Psi(\vec q) \bigg[\varepsilon\!\!\! /_2\, \frac{1}{\not\! p_1-\not\! k_1}\,
    \varepsilon\!\!\! /_1 +
     \varepsilon\!\!\! /_1\, \frac{1}{\not\! p_1-\not\! k_2}\,
    \varepsilon\!\!\! /_2 \bigg]
        \Bigg\}.
\label{eq2}
\end{eqnarray}
Here the relativistic Salpeter wave function $\Psi(\vec q)$
of $0^{-+}$ state with a definite  parity $(-)$ and
charge conjugation $(+)$ can be written as
\cite{cskimwang,changwang}:
\begin{equation}
\Psi(\vec{q})=\left[ {\gamma_{0}}\varphi_1(\vec{q})
+\varphi_2(\vec{q})+\frac{ {\not\! \vec q}\gamma_{0}}{m_1}
\varphi_1(\vec{q})\right]\gamma_{5}\;,
\label{eq3}
\end{equation}
where $\omega_{1}=\sqrt{m_{1}^{2}+{\vec q}^{2}}$ and
$\omega_{2}=\sqrt{m_{2}^{2}+{\vec q}^{2}}$. The wave function
$\varphi_1(\vec{q})$, $\varphi_2(\vec{q})$ and  bound state mass $M$
can be obtained by solving the full Salpeter equation with the
constituent quark mass as input, and they should satisfy the
normalization condition:
\begin{equation}\int\frac{d{\vec{q}}}{(2\pi)^3}
\frac{8\omega_{1}}{m_{1}}\varphi_1 ({\vec{q}})\varphi_2({\vec{q}})
=2M~.\label{eq4}
\end{equation}

Putting wave function $\Psi(\vec{q})$ into the amplitude
Eq. (\ref{eq2}) and performing the trace, the amplitude becomes
\begin{equation}
T=\frac{\sqrt{3}\; (ee_q)^2 \!}{M}
 \int \!\! \frac{d\vec{q}}{(2\pi)^3}4\varphi_1 ({\vec{q}})
 \epsilon_{\mu\nu\alpha\beta} {P^\mu} {\varepsilon_1^\nu}
 {k_1^\alpha}{\varepsilon_2^\beta}\left[
 \frac{1}{(p_1-k_1)^2}+\frac{1}{(p_1-k_2)^2}
\right]~.
\label{eq5}
\end{equation}
With this relativistic amplitude, the two photon decay width can
be written as
   \begin{eqnarray}
   \Gamma(0^{-+} \rightarrow \gamma\gamma)=12\pi \alpha^2
   e_Q^4 M \left\{
    \int \frac{d\vec{q}}{(2\pi)^3} \varphi_1 (\vec{q})
 \left[
 \frac{1}{(p_1-k_1)^2}+\frac{1}{(p_1-k_2)^2} \right]
   \right\}^2~,
   \label{eq6}
   \end{eqnarray}
where $ \alpha=\frac{e^2}{4\pi}$. One can easily  check that in the
non-relativistic limit (by removing the dependence on the relative momentum
$q$) the decay width  depends on the wave function at the origin.

The two gluon decay width of quarkonium can be easily obtained
from the two photon decay width, with a simple replacement in
the photon decay width formula
\begin{equation}
{e_q}^4{\alpha}^2\longrightarrow
\frac{2}{9}{\alpha_s}^2.
\label{eq7}
\end{equation}
That is:
 \begin{eqnarray}
   \Gamma(0^{-+} \rightarrow gg)=\frac{8}{3}\pi \alpha_s^2
  M \left\{
    \int \frac{d\vec{q}}{(2\pi)^3} \varphi_1 (\vec{q})
 \left[
 \frac{1}{(p_1-k_1)^2}+\frac{1}{(p_1-k_2)^2} \right]
   \right\}^2~.
   \label{eq8}
   \end{eqnarray}

\begin{figure}
\begin{picture}(250,130)(200,400)
\put(0,0){\includegraphics{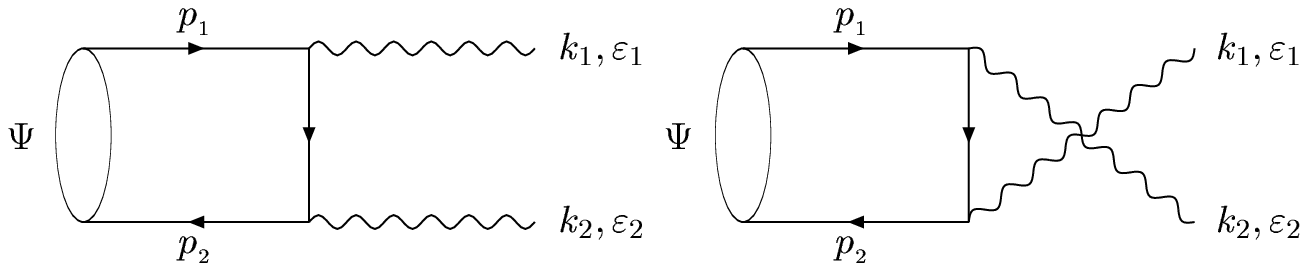}}
\end{picture}
\caption{Two-photon annihilation diagrams of the quarkonium. }
\end{figure}

\section{Numerical Results and Discussions}

In our previous works \cite{cskimwang,changwang}, the full Salpeter
equation has been solved and the corresponding eigenvalue and the wave
function Eq. (\ref{eq3}) have been obtained numerically. We will not
show the details of the calculation here, but only give the
final results; interested readers can find them in Refs.~\cite{cskimwang,changwang}.

When solving the full Salpeter equation, we choose a
phenomenological Cornell potential. There are some parameters in
this potential including the constituent quark mass and one loop
running coupling constant. In previous paper
Ref.~\cite{cskimwang}, the following best-fit values
of input parameters were obtained by fitting the mass spectra
for heavy meson $0^{-}$ states:
\begin{center}
$a=e=2.7183$, $\alpha=0.06$ GeV, $V_0=-0.60$ GeV,
$\lambda=0.2$ GeV$^2$, $\Lambda_{QCD}=0.26$ GeV~~~ and \\
 $m_c=1.7553$ GeV.
\end{center}
With this parameter set, we solve the full Salpeter equation and
obtain the mass spectra and wave functions of quarkonium. We first
fit the mass of $M_{\eta_c}=2.979$ GeV and then get the mass of
${\eta'_c}$: $M_{\eta'_c}=3.566$ GeV, which
is a little lower than the recent experimental data
$M_{\eta'_c}=3.6308\pm 0.0034\pm 0.0010$ GeV \cite{barbar} and
$M_{\eta'_c}=3.654\pm 0.006\pm 0.008$ GeV \cite{belle}. With the
obtained wave function and Eq. (\ref{eq6}), we calculate the decay
width of $\eta_c \rightarrow 2\gamma$ and $\eta'_c \rightarrow 2\gamma$,
with the result:
\begin{equation}
\Gamma(\eta_c \rightarrow \gamma\gamma)=7.14~~{\rm KeV}~,
\end{equation}
\begin{equation}
\Gamma(\eta'_c \rightarrow \gamma\gamma)=4.44~~{\rm KeV}~.
\end{equation}

To give the numerical analysis of two--gluon decays, we need to fix
the value of the renormalization scale $\mu$ in $\alpha_s(\mu)$.
In the case of $\eta_c$ we choose the charm quark mass $m_c$ as
the energy scale and obtain the coupling constant $\alpha_s(m_c)=0.36$
\cite{cskimwang}. The corresponding two--gluon annihilation
rates of $\eta_c$ and $\eta'_c$ are:
\begin{equation}
\Gamma(\eta_c \rightarrow gg)=19.6~~{\rm MeV}~,
\end{equation}
\begin{equation}
\Gamma(\eta'_c \rightarrow gg)=12.1~~{\rm MeV}~.
\end{equation}

For the case of $\eta_b$ and $\eta'_b$, our previously derived input
parameters in the potential should not work, because they were
obtained from fitting data of heavy-light mesons. $\eta_b$ and
$\eta'_b$ being heavy-heavy mesons,  we change the previous scale
parameters to $m_b=5.13$ GeV, $\Lambda_{\rm QCD}=0.2$ GeV, and other
parameters are not changed. With this set of parameters, the mass
of $\eta_b$ is obtained as $M_{\eta_b}=9.364$ GeV, about $100$ MeV
lower than the $\Upsilon$ mass, and  $\eta'_b$ mass as
$M_{\eta'_b}=9.941$ GeV. Now the coupling constant at the scale of
bottom quark mass is $\alpha_s(m_b)=0.232$. The corresponding
decay widths are:
\begin{equation}
\Gamma(\eta_b \rightarrow \gamma\gamma)=0.384~~{\rm KeV}~,
\end{equation}
\begin{equation}
\Gamma(\eta'_b \rightarrow \gamma\gamma)=0.191~~{\rm KeV}~,
\end{equation}
\begin{equation}
\Gamma(\eta_b \rightarrow gg)=6.98~~{\rm MeV}~,
\end{equation}
\begin{equation}
\Gamma(\eta'_b \rightarrow gg)=3.47~~{\rm MeV}~.
\end{equation}

In Table I we list our results with theoretical uncertainties,
which are obtained by varying all the input
parameters simultaneously within $\pm 10\%$ of the central
values, and taking the largest variation of the results. In this
table, we assume the total width of heavy quarkonium is dominated
by its two-gluon decay rate, $\Gamma_{tot}\simeq\Gamma_{2g}$.
The most recent theoretical predictions and
experimental data are also shown in the same table.

\squeezetable
\begin{table*}[hbt]
\setlength{\tabcolsep}{0.5cm} \caption{\small Recent theoretical
and experimental results of two-photon decay width and total
width.} \label{tab1}
\begin{tabular*}{\textwidth}{@{}c@{\extracolsep{\fill}}ccccccc}
 \hline \hline
{\phantom{\Large{l}}}\raisebox{+.2cm}{\phantom{\Large{j}}}
&$\Gamma^{\eta_c}_{2\gamma}$~KeV &$\Gamma^{\eta_c}_{tot}$~MeV&
$\Gamma^{\eta'_c}_{2\gamma}$~KeV&
$\Gamma^{\eta'_c}_{tot}$~MeV&$\Gamma^{\eta_b}_{2\gamma}$~KeV&
$\Gamma^{\eta'_b}_{2\gamma}$~KeV\\ \hline\hline
{\phantom{\Large{l}}}\raisebox{+.2cm}{\phantom{\Large{j}}}
Ours~&~7.14$\pm$0.95 &19.6$\pm$2.6&4.44$\pm$0.48
&12.1$\pm$1.3&0.384$\pm$0.047&0.191$\pm$0.025\\
{\phantom{\Large{l}}}\raisebox{+.2cm}{\phantom{\Large{j}}} M\"unz
\cite{munz} ~&~3.50$\pm$0.40
&&1.38$\pm$0.30&&0.22$\pm$0.04&0.11$\pm$0.02\\
{\phantom{\Large{l}}}\raisebox{+.2cm}{\phantom{\Large{j}}}
Chao \cite{chao}~&~6-7 &17-23&2&5-7&0.46&0.21\\
{\phantom{\Large{l}}}\raisebox{+.2cm}{\phantom{\Large{j}}}
Ebert \cite{ebert} ~&~5.5 &&1.8&&0.35&0.15\\
{\phantom{\Large{l}}}\raisebox{+.2cm}{\phantom{\Large{j}}}
Fabiano \cite{fabiano}~&~7.6$\pm$1.5 &&&&0.466$\pm$101&\\
{\phantom{\Large{l}}}\raisebox{+.2cm}{\phantom{\Large{j}}} Gupta
\cite{gupta} ~&~10.94 &23.03&&&0.46&\\\hline
 {\phantom{\Large{l}}}\raisebox{+.2cm}{\phantom{\Large{j}}}
PDG \cite{pdg}& { $7.2\pm 1.2$}&{ $16.1^{+3.1}_{-2.8}$}&&&&\\
{\phantom{\Large{l}}}\raisebox{+.2cm}{\phantom{\Large{j}}} BABAR
\cite{barbar} & &{ $34.3(2.3)(0.9)$}&&{
$17(8.3)(2.5)$}&&\\
{\phantom{\Large{l}}}\raisebox{+.2cm}{\phantom{\Large{j}}}
BES \cite{bes} & &{ $17(3.7)(7.4)$}&&&&\\
{\phantom{\Large{l}}}\raisebox{+.2cm}{\phantom{\Large{j}}}
CLEO \cite{cleo}& { $7.6(0.8)(2.3)$}&{ $27.0(5.8)(1.4)$}&&&&\\
{\phantom{\Large{l}}}\raisebox{+.2cm}{\phantom{\Large{j}}}
L3 \cite{l3}& { $6.9(1.7)(2.1)$}&&&&&\\
{\phantom{\Large{l}}}\raisebox{+.2cm}{\phantom{\Large{j}}}
AMY \cite{amy}& { $27(16)(10)$}&&&&&\\
{\phantom{\Large{l}}}\raisebox{+.2cm}{\phantom{\Large{j}}}
E760 \cite{e760}& { $6.7^{+2.4}_{-1.7}(2.3)$}&$23.9^{+12.6}_{-7.1}$&&&&\\
 \hline\hline
\end{tabular*}
\end{table*}

{}From the tables, we can see that our results of
$\Gamma^{\eta_c}_{2\gamma}$ agree well with other theoretical estimates of
Refs. \cite{ebert,fabiano,chao}, and $\Gamma^{\eta_c}_{tot}$
with Refs. \cite{chao,gupta}; our results of
$\Gamma^{\eta_b}_{2\gamma}$ and $\Gamma^{\eta'_b}_{2\gamma}$ agree
with Refs. \cite{ebert,fabiano,chao,gupta}; but our results of
$\eta'_c$ are larger than the theoretical predictions by others, but consistent
with the recent experiment data \cite{barbar}.


We comment that in this work we did not include the QCD radiative correction
because we focus mainly on the relativistic corrections,
though there is no doubt that the QCD correction
is very important and an interesting topic.
We have shown the uncertainties of our theoretical estimates by
varying all the input parameters simultaneously within $\pm 10 \%$
of the central values. It should also be pointed out that within these
parameter ranges the uncertainty caused by the value of $\alpha_s(\mu)$
is very important because when we determine the total widths in our
calculation we need the precise value of $\alpha^2_s(\mu)$.

In summary, by solving the relativistic full Salpeter equation
with a well defined form of the wave function, we obtain the mass
spectra: $M_{\eta_c}=2.979\pm 0.432$ GeV, $M_{\eta'_c}=3.566\pm
0.437$ GeV, $M_{\eta_b}=9.364\pm 1.120$ GeV and
$M_{\eta'_b}=9.941\pm 1.112$ GeV. With the help of Mandelstam
formalism for the transition amplitude, we estimate two-photon
decay rates: $\Gamma(\eta_c \rightarrow 2\gamma)=7.14\pm 0.95$
KeV, $\Gamma(\eta'_c \rightarrow 2\gamma)=4.44\pm 0.48$ KeV,
$\Gamma(\eta_b \rightarrow 2\gamma)=0.384\pm 0.047$ KeV and
$\Gamma(\eta'_b \rightarrow 2\gamma)=0.191\pm 0.025$ KeV, and the
total decay widths: $\Gamma_{tot}(\eta_c)=19.6\pm 2.6$ MeV,
$\Gamma_{tot}(\eta'_c)=12.1\pm 1.3$ MeV,
$\Gamma_{tot}(\eta_b)=6.98\pm 0.85$ MeV and
$\Gamma_{tot}(\eta_b)=3.47\pm 0.45$ MeV.\\

\newpage

\acknowledgements

We would like to thank G. Cvetic  for the helpful comments.
The work of C.S.K. was supported in part by  CHEP-SRC
Program and  in part by Grant No. R02-2003-000-10050-0 from BRP of
the KOSEF.
The work of G.W. was supported by BK21 Program, and
in part by Grant No. F01-2004-000-10292-0 of KOSEF-NSFC International
Collaborative Research Grant.
\\


\begin{thebibliography}{99}

\bibitem{barbar}BABAR Collaboration, B. Aubert et al.,
hep-ex/0311038.

\bibitem{pdg} K. Hagiwara $et~ al.$, Particle Data Group,
Phys. Rev. {\bf D66}, 010001 (2002).

\bibitem{belle}BELLE Collaboration, S. K. Choi et al. Phys. Rev.
Lett. {\bf 89}, 102001 (2002).

\bibitem{aleph} A. Heister et al., Phys.
Lett. {\bf B530} 56 (2002).

\bibitem{fang}Fang Shuangshi et al., HEP and NP 27, 277 (2003)(in Chinese).

\bibitem{BS}
E. E. Salpeter and H. A. Bethe, Phys. Rev. {\bf 84}, (1951) 1232.

\bibitem{salp}
E. E. Salpeter, Phys. Rev. {\bf 87}, (1952) 328.


\bibitem{changwang2}Chao-Hsi Chang, Yu-Qi Chen, Guo-Li Wang
and Hong-Shi Zong, Phys. Rev. {\bf D65}, 014017 (2002).

\bibitem{munz} C. M\"unz, {\it Nucl. Phys. A} {\bf 609}, 364 (1996).

\bibitem{mandelstam}S. Mandelstam, Proc. R. Soc. London {\bf 233},
248 (1955).

\bibitem{barbieri}R. Barbieri, R Gatto and R. Kogerler, Phys.
Lett. {\bf B60}, 183 (1976).

\bibitem{keung}W. Y. Keung and I. J. Muzinich, Phys. Rev. {\bf
D27}, 1518 (1983).

\bibitem{chao}Kuang-Ta Chao, Han-Wen Huang, Jing-Hua Liu and Jian
Tang, Phys. Rev. {\bf D56}, 368 (1997).

\bibitem{cskimwang}C. S. Kim and Guo-Li Wang, Phys. Lett. {\bf B584}
(2004) 285;
G. Cvetic, C. S. Kim, Guo-Li Wang and W. Namgung,
Phys. Lett. {\bf B596} (2004) 84.


\bibitem{changwang}
Chao-Hsi Chang, Jiao-Kai Chen  and Guo-Li Wang, hep-th/0312250.

\bibitem{ebert}D. Ebert, R. N. Faustov and V. O. Galkin, Mod. Phy.
Lett. {\bf A18}, 601 (2003).

\bibitem{fabiano}N. Fabiano and G. Pancheri, Eur. Phys. J. {\bf
C25}, 421 (2002).
 N. Fabiano, Eur. Phys. J. {\bf C26}, 441 (2003).

\bibitem{gupta}S. N. Gupta, J. M. Johnson and W. W. Repko, Phys.
Rev. {\bf D54}, 2075 (1996).

\bibitem{bes}BES Collaboration, J. Z. Bai et al., Phys. Lett. {\bf B
555}, 174 (2003).

\bibitem{cleo}CLEO Collaboration, G. Brandenburg et al., Phys.
Rev. Lett. {\bf 85}, 3095 (2000).

\bibitem{l3}L3 Collaboration, M. Acciarri et al., Phys. Lett. {\bf B
461}, 155 (1999).

\bibitem{amy}AMY Collaboration, M. Shirai et al., Phys. Lett. {\bf B
424}, 405 (1998).

\bibitem{e760}E760 Collaboration, T. A. Armstrong et al., Phys. Rev. {\bf
D52}, 4839 (1995).



\end{thebibliography}
\end{document}